\documentclass[reprint,aps,pre,showkeys,showpacs,superscriptaddress,twocolumn,amsmath,amssymb,longbibliography]
{revtex4-1} 
\usepackage{graphics}
\usepackage{bm}
\usepackage{subfigure}
\usepackage{epsfig}
\usepackage[bookmarks,colorlinks,citecolor=blue,linkcolor=blue]{hyperref}
\usepackage{natbib}
\usepackage{siunitx}
\usepackage{soul}
\usepackage{array}
\usepackage{multirow}

\DeclareSIUnit\Molar{\textsc{m}}

\usepackage{xcolor}
\usepackage[normalem]{ulem}

\usepackage{graphicx}	
\begin{document}
\title{Rheology of bidisperse non-Brownian suspensions}
\author{Abhinendra Singh}
\email{abhinendra.singh@case.edu}
\affiliation{Department of Macromolecular Science and Engineering, Case Western Reserve University, Cleveland, OH, 10040, USA}
\affiliation{James Franck Institute, University of Chicago, Chicago, Illinois 60637, USA}
\affiliation{Pritzker School of Molecular Engineering, University of Chicago, Chicago, Illinois 60637, USA}
\author{Christopher Ness}
\affiliation{School of Engineering, University of Edinburgh, Edinburgh EH9 3FG, United Kingdom}
\author{Abhishek K. Sharma}
\affiliation{Pritzker School of Molecular Engineering, University of Chicago, Chicago, Illinois 60637, USA}
\author{Juan J de Pablo}
\affiliation{Pritzker School of Molecular Engineering, University of Chicago, Chicago, Illinois 60637, USA}
\affiliation{Materials Science Division, Argonne National Laboratory, Lemont, Illinois 60439, USA}
\author{Heinrich M Jaeger}
\affiliation{James Franck Institute, University of Chicago, Chicago, Illinois 60637, USA}
\affiliation{Department of Physics, The University of Chicago, Chicago, Illinois 60637, USA}
\date{\today} 
\begin{abstract}
We study the rheology of bidisperse non-Brownian suspensions using particle-based simulation,
mapping the viscosity as a function of the size ratio of the species,
their relative abundance,
and the overall solid content.
The variation of the viscosity with applied stress exhibits shear thickening phenomenology irrespective of composition,
though the stress-dependent limiting solids fraction governing the viscosity and its divergence point are non-monotonic in the mixing ratio.
Contact force data demonstrate an asymmetric exchange in dominant stress contribution from large-large to small-small particle contacts
as the mixing ratio of the species evolves. 
Combining a prior model for shear thickening with one for composition-dependent jamming,
we obtain a full description of the rheology of bidisperse non-Brownian suspensions capable of predicting effects such as the 
viscosity reduction observed upon adding small particle fines to a suspension of large particles.

\end{abstract}
\maketitle
\section{Introduction}
Suspensions of small particles, radius $a\approx$~\SI{100}{\nano\meter}$-$\SI{10}{\micro\meter},
form a class of complex fluids abundant in nature and industry~\cite{Ness_2022, Morris_2020, Singh_2023}.
Their widespread use 
calls for detailed constitutive characterization to enable reliable process design~\cite{roussel2010steady},
especially in the dense regime where particles and fluid are mixed roughly equally~\cite{Brown_2014}.
Under external deformation, these systems, which are apparently simple in composition,
exhibit complex rheology including yielding, shear thinning, shear thickening, and jamming~\cite{Brown_2014, Peters_2016, Singh_2019}.
Recently this phenomenology has been linked to microscopic physics,
specifically constraints that control the relative translation and rotation of interacting particle pairs~\cite{Guy_2018, Singh_2020, Singh_2022}.
Shear thickening,
for example,
represents a crossover from unconstrained to constrained tangential motion as the imposed particle stress $\sigma$ exceeds a threshold set by the repulsive force
$\sigma_0\sim F_0/a^2$~\cite{Seto_2013a}.
A mean-field approach by~\citet{Wyart_2014} (WC) captures the transition
using a stress-dependent jamming volume fraction
$\phi_J (\sigma)$ interpolating between low ($\phi_0$) and high ($\phi_m$) stress limits,
reproducing (in some cases quantitatively) the steady state rheology~\cite{Guy_2018, Singh_2018}.

The above conceptual framework was devised based on nearly monodisperse suspensions, and most numerical and experimental works that seek to test it reflect this~\cite{Singh_2018,Guy_2015,Ness_2016,Guy_2020,Pradeep_2021}. 
As soon as significant deviations from monodispersity are considered,
however,
complexity emerges that is not captured by WC.
In particular, adding {a small quantity of} small particles to a system of large ones can reduce the viscosity under shear~\cite{roussel2010steady, van2018concrete, flatt2004towards},
yet in the reverse case when larger particles are added to small ones, only an increase in viscosity is reported~\cite{cwalina2016rheology, Madraki_2017, Madraki_2018}.
Extending the understanding of constraint-controlled rheology to suspensions with size-disperse particles
is thus a key open problem.
Our approach addresses this challenge in a model system of just two species.
Even in this minimal system, particle size disparity can have a profound effect on the rheology, which, with the exception of a few recent works~\cite{Guy_2020, Malbranche_2023, Pednekar_2018, monti2023shear}, has remained largely unexplored, particularly in the context of the frictional interactions.
{\citet{Pednekar_2018} demonstrated a rheological collapse for polydisperse and bidisperse suspensions once volume fraction $\phi$ is scaled with $\phi_J$, while~\citet{Malbranche_2023} found that the relative viscosity $\eta_r$ for bidisperse suspensions can be well predicted by the usual power law $\eta_r = (1-\phi/\phi_J)^{-2}$ extensively used for quasi-monodisperse dense suspensions~\cite{Singh_2018, Wyart_2014, Guy_2015}.
\citet{Guy_2020} showed that a simple model in which the fraction of frictional contacts enters as a scalar quantity providing a linear interpolation between jamming points fails when one considers the differing stress contributions from different pair classes (large-large, large-small, and small-small). 
Further,~\citet{monti2023shear} linked the reduction in shear thickening behavior of bidisperse suspensions as compared to the monodisperse case to an increase in $\phi_J$ exhibited the suspensions with large size ratio,
an idea we develop in this article.
More recently,~\citet{malbranche2022scaling} extended the universal crossover scaling function for bidisperse suspensions originally proposed based on an experimental dataset by~\citet{Ramaswamy_2023}, demonstrating a collapse for both monodisperse and bidisperse suspensions on the universal curve once scaled with distance from the frictionless jamming for the respective cases.
}

A common feature of the above works is that the rheology of size-disperse suspensions is governed by $\phi_J$,
which
is controlled by (in the bidisperse case) the species size ratio
$\Delta = a_L/a_S$,
their volumetric mixing ratio
$\alpha = N_Sa_S^3/(N_Sa_S^3+N_La_L^3)$,
and the particle friction coefficient $\mu$~\cite{Maranzano_2001}.
{Here $N_L$ and $N_S$ denote the numbers of large and small particles with respective radii being  $a_L$ and $a_S$.}
{\citet{Farris_1968} proposed a model to predict the viscosity of a multimodal suspension based on the idea that the finer particles behave as a liquid with viscosity governed by their volume fraction, which was recently adapted to polydisperse suspensions by~\citet{mwasame2016modeling}.
\citet{Shapiro_1992} used bidisperse glass beads in glycerin and demonstrated that increasing size ratio $\Delta$ leads to higher $\phi_J$. Other studies also showed a similar decrease in viscosity for concentrated bidisperse suspensions as compared to the monodisperse case for a constant $\phi$~\cite{Chong_1971,Barnes_1989, Gondret_1997,Poslinski_1988}.}

Previous experimental~\cite{Gondret_1997,Maranzano_2001, Maranzano_2001a} and numerical~\cite{Chang_1994,Chang_1994a, Chang_1993,Guy_2020, Malbranche_2023, Pednekar_2018} studies show that for a constant solids fraction $\phi$,
the relative suspension viscosity $\eta_r$ decreases with $\Delta$ at fixed $\alpha$ while varying non-monotonically with $\alpha$ at fixed $\Delta$.
This was explained on the basis of $\phi_J$ being non-monotonic in $\alpha$,
for which several models have been proposed~\cite{yu1991estimation,qi2011relative,anzivino2023estimating,servais2002influence,pishvaei2006modelling} {but a complete micromechanical basis is lacking}.
In dry granular materials it is understood that $\phi_J$ depends on $(\Delta,\alpha,\mu)$~\cite{Srivastava_2021},
and there have been attempts to relate this to the fraction of smaller particles being rattlers~\cite{petit2020additional, Srivastava_2021}.

In this article, we systematically explore the role of bidispersity on the rheology of dense, non-Brownian suspensions
using an established simulation scheme~\cite{Seto_2013a, Mari_2015}.
We explore $\Delta \le 12$, $0<\alpha<1$ and $\phi$ close to the jamming point.
{Using simulation data, we provide a micromechanical basis for the well-known non-monotonic dependence $\eta_r(\alpha)$ for a given $\phi$}.
We show that the non-monotonic $\eta_r(\alpha)$ coincides with an exchange of stress contribution dominance from large-large (LL) contacts at small $\alpha$ to small-small (SS) contacts at large $\alpha$,
while the stress carried by large-small (LS) contacts is non-monotonic in $\alpha$ and vanishes at the extremes.
Combining an \emph{ad hoc} extension of {WC with a geometric model for $\phi_J(\Delta,\alpha)$,
we obtain qualitative predictions of the rheology for any bidisperse, shear thickening, non-Brownian suspension.
For simplicity we chose a linear-mixture packing model of the kind described by~\citet{yu1991estimation} (YS),
which takes as its inputs only the size ratio of the species and the pure species jamming points.
Such a model provides the minimal necessary components with which to obtain the nonmonotonicity of $\phi_J$ in species mixing ratio predicted by our simulation model,
but naturally it omits microphysical details such as lubrication, friction and the critical load.
This choice is somewhat arbitrary, though,
since due to the modular combination of models we propose one could in principle introduce here any packing model relevant to a system of interest.
Many such models are present in the literature,
one recent example being by~\citet{anzivino2023estimating}.
}
Using the model, we explore practical settings where small additives are incorporated into suspensions of large particles,
rationalising the counter-intuitive finding that increasing $\phi$ can in some circumstances reduce $\eta_r$.
Meanwhile adding large particles to a suspension of small ones always enhances $\eta_r$, corroborating experimental findings~\cite{Madraki_2017, Madraki_2018}.

\section{Simulation Scheme}
We model $N=6000$ inertialess spheres for $\Delta \le 6$ ($N=12000$ for $\Delta=12$) dispersed in density-matched Newtonian liquid under imposed shear stress $\sigma$ in a constant volume Lees-Edwards periodic domain.
Particles interact through short-range hydrodynamic lubrication and contact forces and torques.
{
Forces on particles obey overdamped dynamics,
governed by a $6N$-dimensional force (and torque) balance between lubrication hydrodynamic ($F_H$) and contact ($F_C$) forces as
\begin{equation}
      \vec{0} = \boldsymbol{F}_{\mathrm{H}}(\boldsymbol{X},\boldsymbol{U}) + \boldsymbol{F}_{\mathrm{C}}(\boldsymbol{X}), 
\end{equation}
where particle positions and velocities velocities are represented by $\boldsymbol{X}$ and $\boldsymbol{U}$, respectively.
In the standard Stokesian Dynamics method~\cite{Brady_1985}, the resistance matrix diverges as $1/h$, where $h$ is the surface separation between particles. In our work, we allow lubrication breakdown~\cite{Ball_1997}, permitting direct contact between particles. To model the direct contacts between particles, we follow \citet{Cundall_1979} and the algorithm by Luding~\cite{Luding_2008}. We make use of the lubrication resistance~\cite{Mari_2015}, and hence do not use dashpot explicitly. The tangential $\boldsymbol{F}^t_{\mathrm{C}}$ and normal contact $\boldsymbol{F}^n_{\mathrm{C}}$ forces between particles satisfy the Coulomb criterion, $|\boldsymbol{F}^t_{\mathrm{C}}| \le \mu|\boldsymbol{F}^n_{\mathrm{C}}|$, for compressive normal forces.
Here, we use $\mu=1$, which has been shown to yield quantitative comparison with experimentally observed rheology~\cite{Mari_2015, Singh_2020, Singh_2022}. Rate dependence is introduced using the so-called Critical Load Model (CLM)~\cite{Seto_2013a, Mari_2014}, where frictional force is activated above a threshold normal force $F_0$ giving a characteristic stress scale $\sigma_0 = F_0/a^2$ (here $a$ is the particle size in the monodisperse limit, and we follow~\citet{Guy_2015, Guy_2020} and assume $F_0$ to be independent of particle size so that $\sigma_0 = F_0/a_s^2$ throughout). Such a stress scale originates from an electrostatic double-layer interaction between particles from the polymer coating, as an example. The CLM model can be considered a special case of debye length $\lambda$ approaching zero.
}

Under imposed constant shear stress $\sigma$, the suspension flows with time-dependent shear rate $\dot{\gamma}(t)$, and we compute the relative viscosity as $\eta_r(t) = \sigma/\eta_0\dot{\gamma}(t)$,
with $\eta_0$ the liquid viscosity.
Rheology data shown in the following are averages of $\eta_r$ over 5 strain units at steady state,
across 5 realizations. We perform simulations for $\Delta  = 2,3,4,6,12$ and $\alpha = [0.05,0.9]$.

%
%

\begin{figure*}[]
\includegraphics[trim = 0mm 65mm 0mm 0mm, clip,width=0.9\textwidth,page=1]{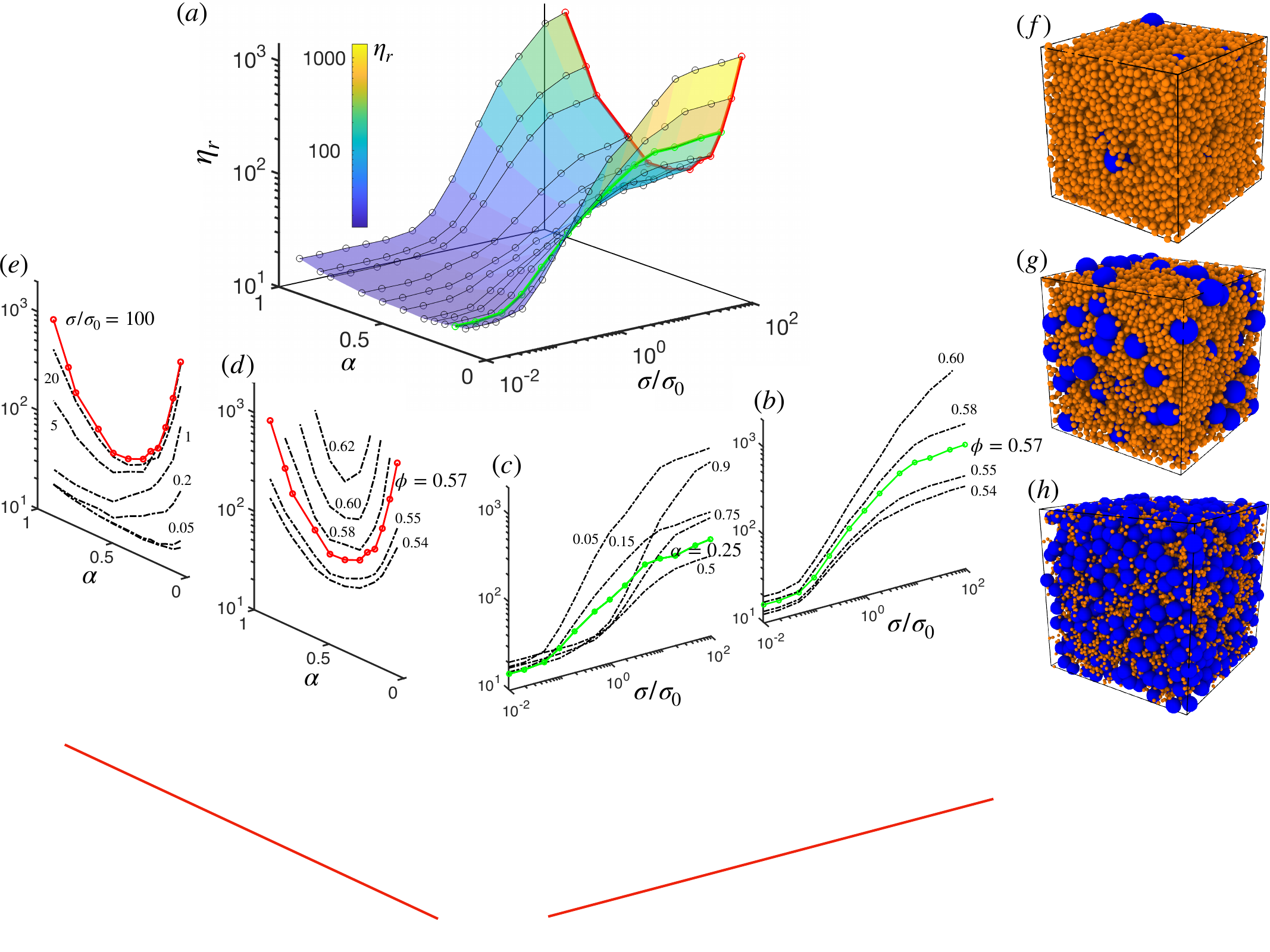}
\caption{
Effect of mixing small and large particles on the rheology of a model suspension
with size ratio $\Delta=4$.
Shown are
(a) the relative viscosity $\eta_r$ as a function of the volume ratio of small particles $\alpha$ and the applied stress $\sigma/\sigma_0$, at $\phi=0.57$.
Red and green lines represent common data sets across each panel, showing,
respectively,
$\eta_r(\alpha)$ for $\sigma/\sigma_0=100$, and $\eta_r(\sigma)$ for $\alpha=0.25$.
(b) $\eta_r$ as a function of $\sigma/\sigma_0$ at fixed $\alpha=0.25$ for various $\phi$.
(c) $\eta_r$ as a function of $\sigma/\sigma_0$ at fixed $\phi=0.57$ for various $\alpha$.
(d) $\eta_r$ as a function of $\alpha$ for various $\phi$ at fixed $\sigma/\sigma_0=100$
(e) $\eta_r$ as a function of $\alpha$ for $\phi=0.57$ for various $\sigma/\sigma_0$.
(f)-(h) Snapshots of the simulation at $\phi=0.57, \sigma/\sigma_0=100$, $\Delta=4$, and (top-to-bottom) $\alpha=0.9,0.5,0.1$.
}
\label{figure1}
\end{figure*}

\section{Results}

\paragraph*{Bidisperse suspension rheology.}
Figure~\ref{figure1} shows the effect of $\alpha$ on the rheology for exemplar data with $\Delta =4$.
We present a map of $(\eta_r, \alpha, \sigma)$ at $\phi=0.57$ in Fig.~\ref{figure1}(a) showing an overview of the behavior,
with planar slices showing
(b) $\eta_r(\sigma/\sigma_0,\phi)$ at $\alpha=0.25$;
(c) $\eta_r(\sigma/\sigma_0,\alpha)$ at $\phi=0.57$;
(d) $\eta_r(\alpha,\phi)$ at $\sigma/\sigma_0=100$;
(e) $\eta_r(\alpha,\sigma/\sigma_0)$ at $\phi=0.57$.
In Figs.~\ref{figure1}(f)-(h) are snapshots of the simulation at $\alpha=0.9,0.5,0.1$.
In Fig~\ref{figure1}(b), one observes canonical thickening behavior, qualitatively similar to quasi-monodisperse systems showing a stress-mediated transition between two Newtonian plateaus {driven by activation of frictional contacts}~\cite{Mari_2014, Guy_2015, Ness_2016}. 
The viscosities of the plateaus increase with $\phi$ towards their respective jamming points,i.e., {frictionless
$\phi_J^0\equiv\phi_J(\sigma/\sigma_0\to0)\approx0.75$ and frictional $\phi_J^\mu\equiv\phi_J(\sigma/\sigma_0\to\infty)\approx0.65$  limits
(these numbers being sensitive to $\alpha$, $\Delta$ and $\mu$).}
In Fig.~\ref{figure1}(c) we find that $\sigma_0$ increases monotonically with $\alpha$ (since the former is related to the particle size through $\sigma_0\sim F_0/a^2$),
while the frictionless and frictional viscosities measured at $\sigma/\sigma_0=0.01$ and $\sigma/\sigma_0=100$ respectively show non-monotonic dependence on $\alpha$.
The crossover of these flow curves is a result of the combined effect of bidispersity on $\sigma_0$ and $\phi_J$ and is not predicted by models that assume monodispersity.
Next, we present data in the $\eta_r(\alpha)$ plane.
Figure~\ref{figure1}(d) shows $\eta_r$ for $\sigma/\sigma_0=100$ as a function of $\alpha$ for various $\phi$.
At fixed $\phi$,
$\eta_r$ first decreases with increasing $\alpha$,
reaching a minimum before increasing again so that $\eta_r(\alpha=0) = \eta_r(\alpha=1)$.
In the limits $\alpha = (0,1)$ the suspension is monodisperse and exhibits identical rheology due to the size invariance of non-Brownian suspensions (when $\sigma/\sigma_0$ is 0 or $\infty$).
The value of $\alpha$ that minimizes $\eta_r$ is insensitive to $\phi$.
{
The mentioned behavior is consistent with the literature on flowing suspensions~\cite{Gondret_1997,Maranzano_2001, Maranzano_2001a,Chang_1994,Chang_1994a, Chang_1993,Guy_2020, Malbranche_2023, Pednekar_2018}. However, none of the studies explored jammed states.}
For $\phi \ge 0.58$, the suspension near the extrema $\alpha=0,1$ is jammed,
and
the window of $\alpha$ for which $\eta_r$ is finite decreases with increasing $\phi$ so that no flow occurs at any $\alpha$ for $\phi \ge 0.65$.
Figure~\ref{figure1}(e) shows $\eta_r(\alpha)$ for various $\sigma/\sigma_0$ at fixed $\phi=0.57$.
For $\sigma/\sigma_0=100$, $\eta_r$ at the large and small limits of $\alpha$ are equal,
while for $\sigma/\sigma_0 \le 5$,
$\eta_r$ is higher for small $\alpha$ ($\alpha \to 0$) as compared to larger values ($\alpha \to 1$).
This can be explained based on the relation between $\sigma_0$ and particle size $a$. Since $\sigma_0\sim1/a^2$, $\sigma_0$ increases with $\alpha$ (Fig.~\ref{figure1}(c)) so that the jamming point is governed by friction at lower stress when particles are large.

\begin{figure*}
\includegraphics[trim = 0mm 85mm 0mm 0mm, clip,width=0.9\textwidth,page=2]{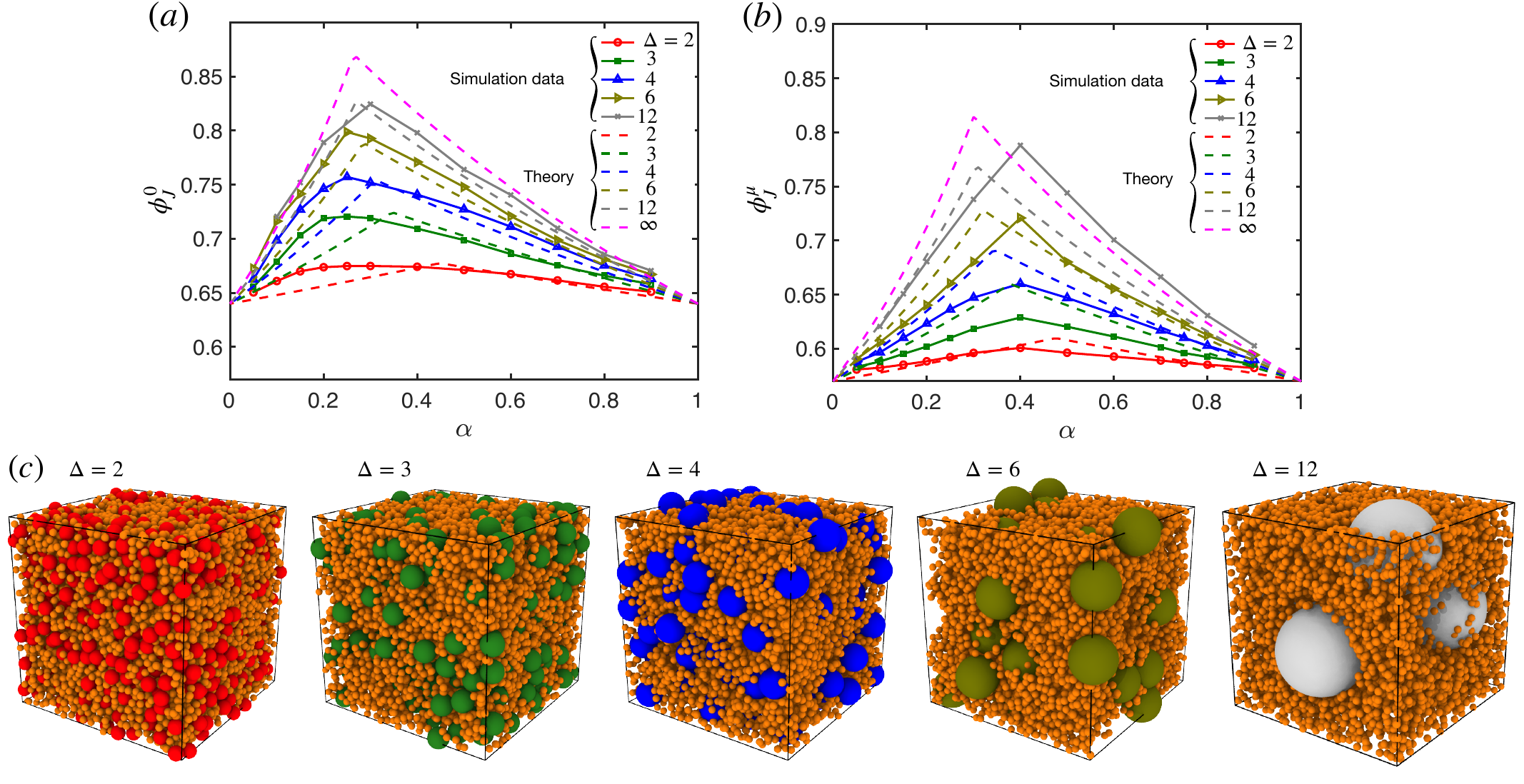}
\caption{
Role of the particle size ratio $\Delta$.
(a) Variation with $\alpha$ of $\phi_J^0$,
the jamming point at $\sigma/\sigma_0 =0$,
for a range of $\Delta$.
Solid lines with points represent simulation data;
dashed lines show model predictions~\cite{yu1991estimation};
(b) Variation with $\alpha$ of $\phi_J^\mu$,
the jamming point at $\sigma/\sigma_0 =\infty$,
for a range of $\Delta$.
(c) Snapshots of simulations with $\phi=0.5$, $\mu=1$, $\alpha=0.5$ and (left to right) $\Delta=2$, $3$, $4$, $6$, $12$.
}
\label{figure2}
\end{figure*}

Figure~\ref{figure2} shows the effect of $\Delta$ on the jamming points $\phi_J^0$, $\phi_J^\mu$.
In Figs.~\ref{figure2}(a) and (b) are the variation of $\phi_J^0$ and $\phi_J^\mu$ with $\alpha$, for $\Delta = (2,3,4,6,12)$, obtained by simulating the limits $\sigma/\sigma_0 \to 0$ (lubricated, frictionless state) and $\sigma/\sigma_0 \to \infty$ (frictional state) for a range of $\phi$, then fitting the viscosity as $\eta_r = (1-\phi/\phi_{J}^{\{0,\mu\}})^{-2}$~\cite{krieger1959mechanism}.
Both $\phi_J^0$ and $\phi_J^{\mu}$ vary non-monotonically with $\alpha$,
with the dependence being more pronounced for increasing $\Delta$, {which is consistent with previous experimental and numerical findings}.
Also plotted are $\phi_{J}^{\{0,\mu\}}$ predictions as functions of $\alpha$ at various $\Delta$ following the model of~\citet{yu1991estimation},
which produces jamming point predictions based on geometry only.
Interestingly the model works better for the frictionless limit
($\phi_J^0$, $\sigma/\sigma_0=0$)
as compared to the frictional one
($\phi_J^\mu$, $\sigma/\sigma_0=\infty$),
likely due to the absence of friction and shear-induced structure in the theory.
(Indeed, understanding the disparity in the nature of jamming between frictionless and frictional systems remains an open challenge~\cite{liu2010jamming}.)
{
In the frictionless case, the simulation data appear to be converging toward the theory for larger values of $\alpha>0.4$. Though for smaller values of $\alpha$, discrepancies can be observed. It is also important to mention that testing this for larger $\Delta$ rapidly becomes computationally intractable.}
Plotting $\eta_r$ as a function of $\sigma/\sigma_0$ for different $\Delta$ at fixed $\alpha=0.5, \phi=0.57$ (Fig.~\ref{figure2}(c)),
one finds the viscosity of both limiting states decreases with increasing $\Delta$, as the proximity to jamming is decreased.

\begin{figure}[b]
\includegraphics[trim = 0mm 18mm 0mm 0mm, clip,width=0.475\textwidth,page=4]{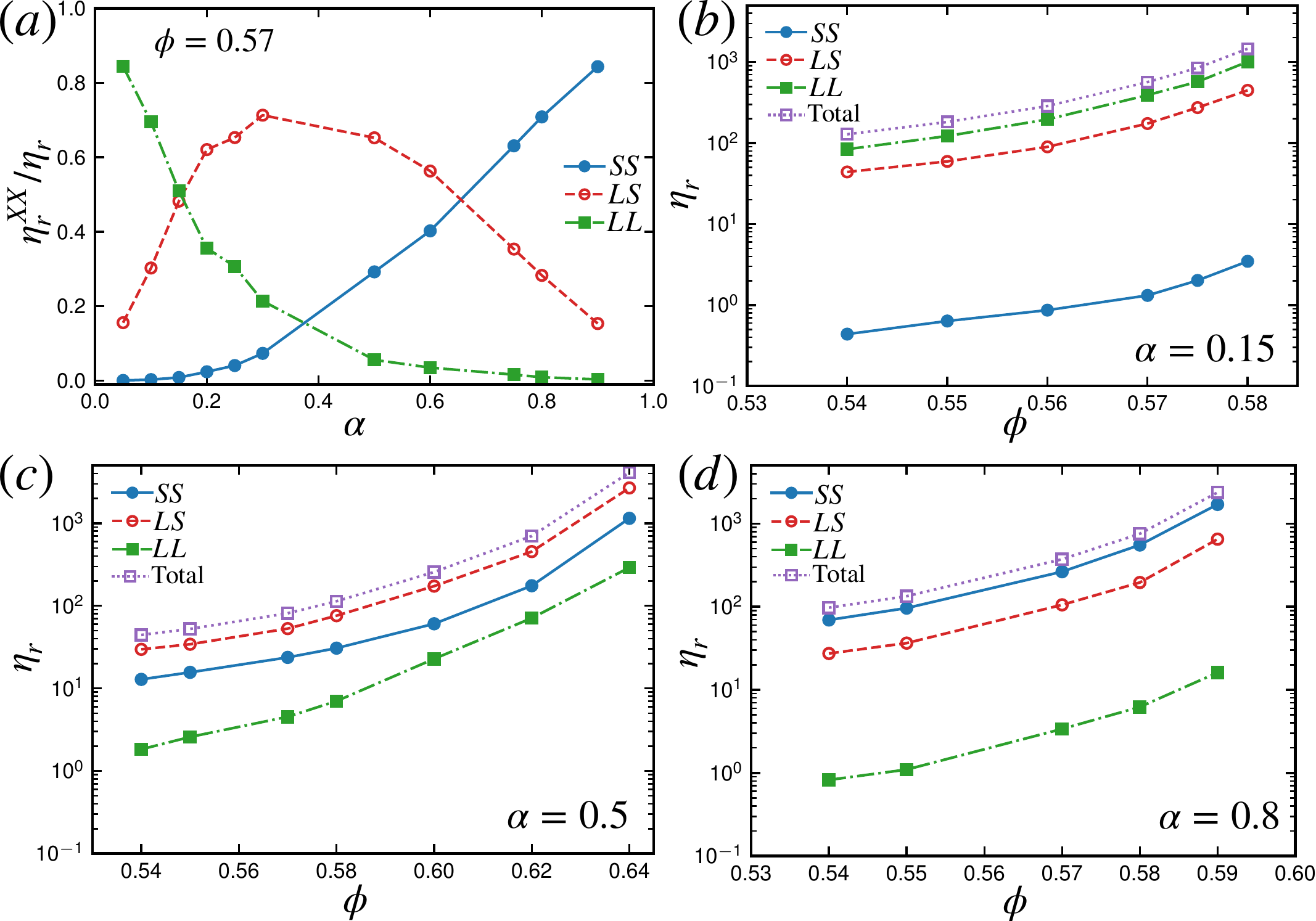}
\caption{
Viscosity contributions from each contact type.
(a) Relative contribution $\eta_r^{XX}$ to the total viscosity $\eta_r$ of each contact type (where SS, LS, and LL replace $XX$) as a function of $\alpha$,
for $\phi=0.57$ and $\sigma/\sigma_0=100$; 
(b)-(d) Total viscosity and its contact contributions as a function of $\phi$ at $\sigma/\sigma_0=100$ for $\alpha=$
(b) 0.15;
(c) 0.5, and 
(d) 0.8;
}
\label{figure4}
\end{figure}

\begin{figure*}
\includegraphics[trim = 0mm 0mm 160mm 0mm, clip,width=0.6\textwidth,page=3]{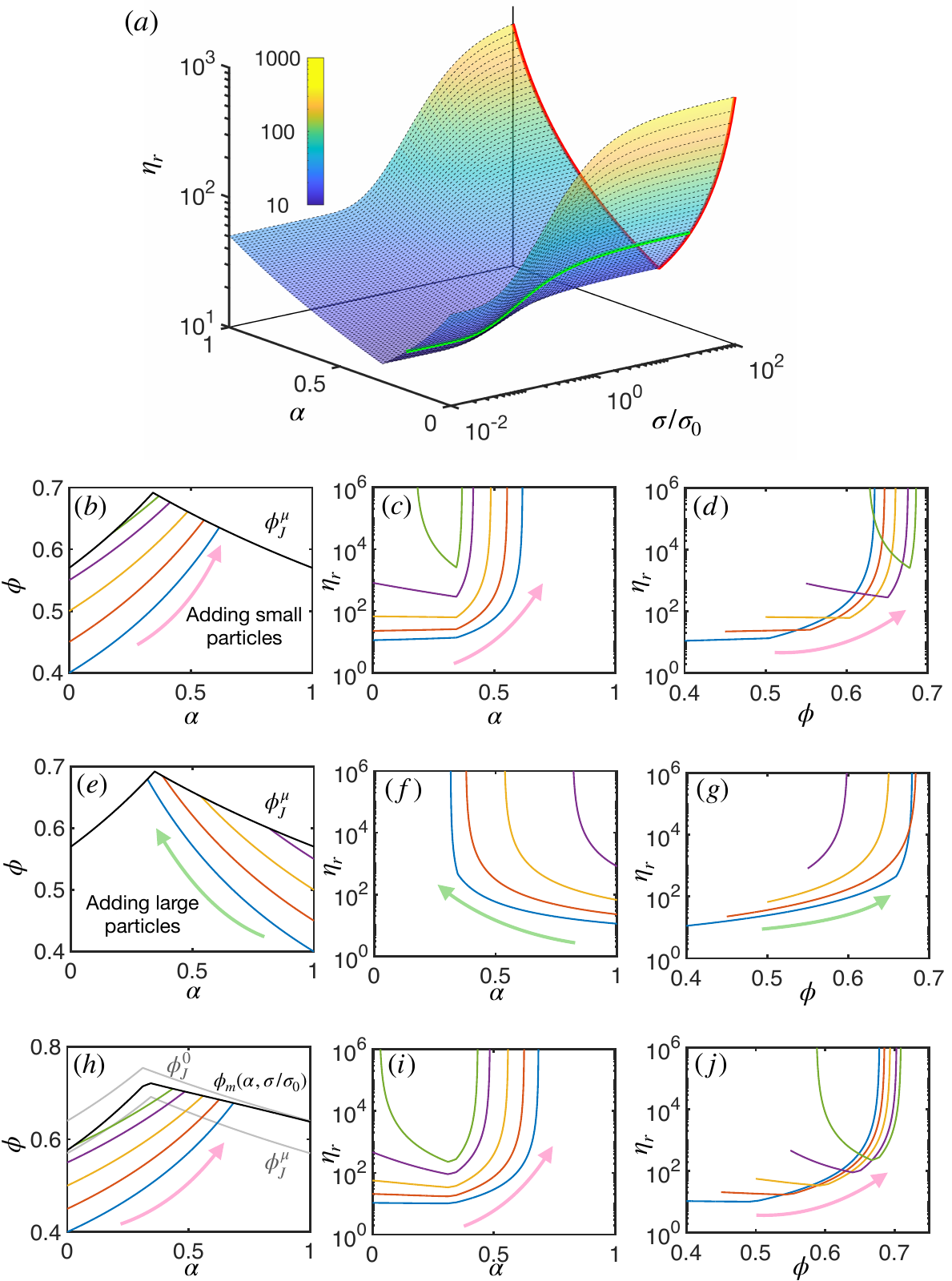}
\caption{
Constitutive model predictions.
(a) Combining the WC model for $\eta_r(\sigma/\sigma_0)$ with the YS model for $\phi_J(\alpha,\Delta)$, one obtains the rheology of bidisperse frictional suspensions.
{
(b)-(j) Colors represent different initial volume fraction $\phi$ with blue representing the lowest inital $\phi$ and black lines represent jamming volume fraction $\phi_J^\mu(\alpha)$ and $\phi_m(\alpha,\sigma/\sigma_0$).
Colored arrows indicate adding small (pink) and large (green) particles.
}
(b)-(d) Adding small particles to a monodisperse large particle packing at $\sigma/\sigma_0=\infty$:
(b) Variation of $\phi$ and $\phi_m$ with $\alpha$, and $\eta_r$ plotted as a function of (c) $\alpha$ and (d) $\phi$. 
(e)-(g) Adding large particles to a monodisperse small particle packing at $\sigma/\sigma_0=\infty$:
(e) Variation of $\phi$ and $\phi_m$ with $\alpha$, and $\eta_r$ plotted as a function of (f) $\alpha$ and (g) $\phi$.
(h)-(i) Adding small particles to a monodisperse large particle packing at intermediate $\sigma/\sigma_0$, so that large particles are frictional and small ones are frictionless.
(h) Variation of $\phi$ and $\phi_J$ with $\alpha$, and $\eta_r$ plotted as a function of (i) $\alpha$ and (j) $\phi$.
Also shown in (h) in gray are $\phi_m$ and $\phi_0$.
}
\label{figure3}
\end{figure*}

\paragraph*{Contribution of different contact types.}
We next address the microscopic underpinning of the non-monotonic $\eta_r(\alpha)$ reported in Fig.~\ref{figure1} (see also~\cite{Gondret_1997,Maranzano_2001, Maranzano_2001a,Chang_1994,Chang_1994a, Chang_1993,Guy_2020, Malbranche_2023, Pednekar_2018}). {In the literature, this behavior was explained based on the non-monotonic behavior of $\phi_J(\alpha)$, making the distance to jamming $(\phi_J-\phi)$ and thus $\eta_r$ to be non-monotonic at a given constant $\phi$.}
We use simulation data to separate the stress contributions of different types of contacts LL {(large-large)}, LS {(large-small)}, and SS {(small-small)}.
Figure~\ref{figure4}(a) shows the viscosity contribution of each contact type scaled with $\eta_r$ for $\phi=0.57$ at $\sigma/\sigma_0=100$ as a function of $\alpha$ for $\Delta=4$.
{In the limits $\alpha = \{0,1\}$, $\eta_r$ would be purely dominated by large-large (LL) and small-small (SS) contacts with LS contribution being zero in the two limits.}
At small $\alpha$, large-large (LL) contacts provide the dominant contribution to $\eta_r$, while SS contacts take over at large $\alpha$ as expected.
LL decay from 1 at $\alpha=0$ (by definition) to $\approx0$ for $\alpha \ge 0.6$.
SS contributions are minimal for $\alpha \le 0.2$ and increasing to 1 at $\alpha=1$.
Notably, the value of $\alpha$ at which SS contributions begin to increase and LL vanishes are not symmetric with respect to 0 and 1.
The contribution of LS is non-monotonic and vanishes for $\alpha=0,1$, and is maximal around $\alpha =0.4$ where it contributes $\approx~$75\% of the overall viscosity.
{
LS contribution is the dominant viscosity component for $\alpha \in \{0.2, 0.75\}$.}
Generalizing these findings for different values of $\phi$, 
Figs.~\ref{figure4}(b)-(d) present $\eta_r$ as a function of $\phi$ for $\alpha=$ 0.15, 0.5, and 0.8 respectively.
{We observe that for small $\alpha=0.15$, the dominant contribution to viscosity originates from LL contacts. The viscosity of large-large particle contacts $\eta_r^{LL}$ is nearly 2-3 orders of magnitude larger compared to $\eta_r^{SS}$, while being only 2-3 times larger than $\eta_r^{LS}$. On the other hand, for $\alpha=0.8$, $\eta_r^{SS}$ is 2-3 orders of magnitude larger than $\eta_r^{LL}$. Hence, in the two extreme limits of $\alpha$, jamming originates from the dominant contributions from LL and SS components, with LS contributions being the second dominant one. LS provides the major contribution to viscosity for the intermediate case $(\alpha=0.5)$, followed by SS and LL being the least significant contributor to viscosity.
}

\paragraph*{Constitutive model.}
{\citet{Guy_2020} demonstrated that WC model and its variants~\cite{Wyart_2014, Singh_2018} as postulated fail to reproduce the bidisperse rheology.
}
Given the qualitative agreement of the dependence of $\phi_J^0$ and $\phi_J^\mu$ on $\alpha$ and $\Delta$ with YS~(Fig.~\ref{figure2}) and of $\eta_r(\sigma/\sigma_0)$ with WC in our previous nearly monodisperse cases~\cite{Singh_2018, Singh_2020}, we are motivated to construct a combined model to capture the full behavior.

{
Conventional WC assumes that the suspension viscosity diverges at stress-dependent jamming volume fraction as 
\begin{equation}
 \eta_r(\phi,\sigma/\sigma_0) = (1-\phi/\phi_m(\sigma/\sigma_0))^{-2},
\end{equation}\label{eq:eta_sigma}
where $\phi_J^0$ and $\phi_J^\mu$ denote the frictionless and frictional jamming volume fractions. The stress-dependent jamming volume fraction $\phi_m(\sigma/\sigma_0)$ is postulated as
\begin{equation}
   \phi_m(\sigma/\sigma_0) = f(\sigma/\sigma_0)\phi_J^\mu + (1-f(\sigma/\sigma_0))\phi_J^0,
\end{equation}
with the fraction of frictional contacts being expressed as 
$f(\sigma/\sigma_0) = \exp(-\sigma_0/\sigma)$. Here $\sigma_0$ is the onset stress for thickening.
}

{
To extend the WC model to bidisperse systems, we let $\phi_J^0$ and $\phi_J^\mu$ be functions of $\alpha$ and $\Delta$ as plotted in Fig.~\ref{figure2}(a)-(b). To incorporate particle size dependence, $\sigma_0$ is interpolated in an \emph{ad hoc} way between small and large particle limits as $\sigma_0(\alpha) = F_0/(\alpha a_S + (1-\alpha)a_L)^2$. Substituting $\sigma_0(\alpha)$, $\phi_J^0(\Delta,\alpha)$, and $\phi_J^\mu(\Delta,\alpha)$ into Eq.~\eqref{eq:eta_sigma}, we propose
\begin{equation}
    \eta_r(\phi,\sigma/\sigma_0,\Delta,\alpha) = (1-\phi/\phi_m(\sigma/\sigma_0,\Delta,\alpha))^{-2}.
\end{equation}\label{eq:eta_sigma_bidis}
We note that plotting three-dimensional data together with the model becomes difficult to interpret, hence in the following we only discuss results from the constitutive model, where the simulations are in close agreement. }

%
Using this model we can predict the effect on the rheology of compositional changes.
{As small (or large) particles are added to an initial suspension of large (or small) particles, $\alpha$ increases (or decreases) accompanied by an increase in $\phi$.}
First, we consider in Fig.~\ref{figure3}(b)-(d) the effect of adding small particles to a suspension of monodisperse large ones, for $\Delta=4$ in the frictional limit ($\sigma/\sigma_0=\infty$). Here, colors indicate increasing initial volume fraction $\phi$ from blue to black.
{
Both $\alpha$ and $\phi$ increase with the number of small particles. In the case of dense suspensions, $\eta_r$ ideally increases with $\phi$, eventually diverging at the jamming volume fraction~\cite{Morris_2020}.}
Counter-intuitively, for sufficiently large initial $\phi$ (of large particles),
the addition of small particles leads to a decrease in $\eta_r$.
Even though both $\phi$ and $\alpha$ increase as small particles are added, the model predicts that the increase in $\phi$ is slower than that of $\phi_m$, hence $(1-\phi/\phi_m)^{-2}$ decreases.
Conversely, the asymmetry in $\phi_m(\alpha)$ means adding large particles to a packing of small ones leads, more intuitively, to an increase in both $\phi$ and $\eta_r$ (Fig.~\ref{figure3}(e)-(g)).
At intermediate $\sigma/\sigma_0$, $\phi_J$ interpolates between frictionless and frictional values (Fig.~\ref{figure3}(h)),
and the range of $\alpha$ for which viscosity reduction appears is broadened (Fig.~\ref{figure3}(i)-(j)).
Recent experiments have shown that adding large non-Brownian particles to a suspension of small ones leads to enhanced thickening behavior~\cite{Madraki_2018, Madraki_2017}, consistent with our prediction (Fig.~\ref{figure3}(e)) that doing so will always move the system closer to its respective $\phi_m$.

\section{Concluding remarks}
{
In this work, we have studied the rheology of dense bidisperse suspensions by extensive numerical simulations. The simulations presented here consider lubrication hydrodynamics interactions and repulsive normal contact forces, where static frictional force is activated beyond a critical threshold. The rheological flow curves display rate-independent rheology at small stresses $(\sigma \ll \sigma_0)$ followed by thickening behavior. The shear thickening behavior presented depends on the overall volume fraction $\phi$ as well as packing properties ($\Delta,\alpha$).
}

{
For a given large enough $\phi$ and $\Delta$, discontinuous shear thickening (DST) is observed around the extreme values of $\alpha$, while continuous shear thickening for intermediate $\alpha$.
Viscosity $\eta_r$ shows a non-monotonic dependence as a function of $\alpha$ for a constant ($\phi,\sigma/\sigma_0$). 
We address this behavior by separating the stress contributions from different types of contacts, viz., large-large (LL), large-small (LS), and small-small (SS) contacts. We find that at the extreme values of $\alpha$, viscosity is dominated by LL and SS contributions with LS contribution going to zero (by definition).
SS contribution goes much more rapidly as compared to the decrease in SS contribution.
LS contribution is non-monotonic and carries the majority of stress for intermediate $\alpha \in \{0.2, 0.7\}$ values. 
}

%
We have shown that an existing constitutive relation for rate-dependent rheology (WC,~\cite{Wyart_2014}) can be combined in an \emph{ad hoc} way with a geometric model by~\citet{yu1991estimation} to obtain a qualitative picture of the rheology of bidisperse suspensions.
{We emphasize that using the geometric model by~\citet{yu1991estimation} is purely a choice, many other models~\cite{anzivino2023estimating,qi2011relative} should yield similar results, perhaps yielding better qualitative predictions for specific systems.}
Our ansatz for $\sigma_0(\alpha)$ is not strictly in
accordance with WC, in the sense that the order parameter f can no longer represent the fraction of contacts that
are frictional (see Ref~\cite{Guy_2020} for more details), { and we did not explicitly explore different types of frictional contacts (LL,LS, and SS types). }
%
%
Nonetheless, our model predicts novelty in the rheology that is present in reality but unaddressed by WC,
specifically 
that adding small particles to a system of large ones can counterintuitively decrease the viscosity,
whereas in the opposite case adding large particles to small particles always leads to an increase in $\eta_r$.  
We show additionally that the non-monotonic dependence of relative viscosity $\eta_r$ on $\alpha$ for constant $\phi$ (as reported by~\cite{Maranzano_2001, Maranzano_2001a, Pednekar_2018, Malbranche_2023}) can be understood by delineating the stress contributions of each type of contact,
and our results suggest that the rheology at $\Delta=12$ is already close to the $\Delta\to\infty$ limit, so the predictions made here will be useful across bidisperse systems of arbitrary size ratio.
The consequences of these results for more complex systems, especially polydisperse samples~\cite{mwasame2016modeling} in which colloidal forces may become relevant~\cite{li2023simulating}, are broad.
In particular, our results provide a direction towards the limiting case of predicting the rheology of mixtures where larger additives are included in a continuous background phase of much smaller particles~\cite{cwalina2016rheology}.
%
{
Recent work by Pednekar et al.~\cite{Pednekar_2018} demonstrated that the rheology of a polydisperse suspension (viscosity, normal stresses, and particle pressure) can be represented by an equivalent bidisperse suspension. This work was in parts inspired by previous works by Desmond \& Weeks\cite{Desmond_2014}, Ogarko \& Luding~\cite{Ogarko_2012} in dry granular jamming showing equivalence in jamming behavior between polydisperse and equivalent bidisperse systems.}
This suggests that our results can similarly be used to
make predictions for polydisperse suspensions, which
has many applications:
industrial processing of slurries and muds;
manufacturing of amorphous solid dispersions;
and predicting the runout of geophysical flows comprising
grains spanning many orders of magnitude~\cite{breard2023fragmentation}.

\paragraph*{Acknowledgements}
Codes and scripts necessary to reproduce the results reported in this article are available on request.
AS and HMJ acknowledge support from the Army Research Office under grant W911NF-21-2-0146.
A. S. acknowledges Case Western Reserve University for start-up funding.
C.N. thanks Eric Breard for pointing to Ref~\cite{yu1991estimation}, and acknowledges support from the
Royal Academy of Engineering under the Research Fellowship scheme
and from the Leverhulme Trust under Research Project Grant RPG-2022-095.
Part of this work made use of the High Performance Computing Resource in the Core Facility for Advanced Research Computing at Case Western Reserve University.

\vspace{2mm}
A.S. and C.N. contributed equally to this work.
\bibliography{dst}

\begin{thebibliography}{58}%
\makeatletter
\providecommand \@ifxundefined [1]{%
 \@ifx{#1\undefined}
}%
\providecommand \@ifnum [1]{%
 \ifnum #1\expandafter \@firstoftwo
 \else \expandafter \@secondoftwo
 \fi
}%
\providecommand \@ifx [1]{%
 \ifx #1\expandafter \@firstoftwo
 \else \expandafter \@secondoftwo
 \fi
}%
\providecommand \natexlab [1]{#1}%
\providecommand \enquote  [1]{``#1''}%
\providecommand \bibnamefont  [1]{#1}%
\providecommand \bibfnamefont [1]{#1}%
\providecommand \citenamefont [1]{#1}%
\providecommand \href@noop [0]{\@secondoftwo}%
\providecommand \href [0]{\begingroup \@sanitize@url \@href}%
\providecommand \@href[1]{\@@startlink{#1}\@@href}%
\providecommand \@@href[1]{\endgroup#1\@@endlink}%
\providecommand \@sanitize@url [0]{\catcode `\\12\catcode `\$12\catcode
  `\&12\catcode `\#12\catcode `\^12\catcode `\_12\catcode `\%12\relax}%
\providecommand \@@startlink[1]{}%
\providecommand \@@endlink[0]{}%
\providecommand \url  [0]{\begingroup\@sanitize@url \@url }%
\providecommand \@url [1]{\endgroup\@href {#1}{\urlprefix }}%
\providecommand \urlprefix  [0]{URL }%
\providecommand \Eprint [0]{\href }%
\providecommand \doibase [0]{http://dx.doi.org/}%
\providecommand \selectlanguage [0]{\@gobble}%
\providecommand \bibinfo  [0]{\@secondoftwo}%
\providecommand \bibfield  [0]{\@secondoftwo}%
\providecommand \translation [1]{[#1]}%
\providecommand \BibitemOpen [0]{}%
\providecommand \bibitemStop [0]{}%
\providecommand \bibitemNoStop [0]{.\EOS\space}%
\providecommand \EOS [0]{\spacefactor3000\relax}%
\providecommand \BibitemShut  [1]{\csname bibitem#1\endcsname}%
\let\auto@bib@innerbib\@empty
\bibitem [{\citenamefont {Ness}\ \emph {et~al.}(2022)\citenamefont {Ness},
  \citenamefont {Seto},\ and\ \citenamefont {Mari}}]{Ness_2022}%
  \BibitemOpen
  \bibfield  {author} {\bibinfo {author} {\bibfnamefont {C.}~\bibnamefont
  {Ness}}, \bibinfo {author} {\bibfnamefont {R.}~\bibnamefont {Seto}}, \ and\
  \bibinfo {author} {\bibfnamefont {R.}~\bibnamefont {Mari}},\ }\href@noop {}
  {\bibfield  {journal} {\bibinfo  {journal} {Annual Review of Condensed Matter
  Physics}\ }\textbf {\bibinfo {volume} {13}},\ \bibinfo {pages} {97} (\bibinfo
  {year} {2022})}\BibitemShut {NoStop}%
\bibitem [{\citenamefont {Morris}(2020)}]{Morris_2020}%
  \BibitemOpen
  \bibfield  {author} {\bibinfo {author} {\bibfnamefont {J.~F.}\ \bibnamefont
  {Morris}},\ }\href@noop {} {\bibfield  {journal} {\bibinfo  {journal} {Annual
  Review of Fluid Mechanics}\ }\textbf {\bibinfo {volume} {52}},\ \bibinfo
  {pages} {121} (\bibinfo {year} {2020})}\BibitemShut {NoStop}%
\bibitem [{\citenamefont {Singh}(2023)}]{Singh_2023}%
  \BibitemOpen
  \bibfield  {author} {\bibinfo {author} {\bibfnamefont {A.}~\bibnamefont
  {Singh}},\ }\href@noop {} {\bibfield  {journal} {\bibinfo  {journal} {MRS
  Communications}\ ,\ \bibinfo {pages} {1}} (\bibinfo {year}
  {2023})}\BibitemShut {NoStop}%
\bibitem [{\citenamefont {Roussel}\ \emph {et~al.}(2010)\citenamefont
  {Roussel}, \citenamefont {Lema{\^\i}tre}, \citenamefont {Flatt},\ and\
  \citenamefont {Coussot}}]{roussel2010steady}%
  \BibitemOpen
  \bibfield  {author} {\bibinfo {author} {\bibfnamefont {N.}~\bibnamefont
  {Roussel}}, \bibinfo {author} {\bibfnamefont {A.}~\bibnamefont
  {Lema{\^\i}tre}}, \bibinfo {author} {\bibfnamefont {R.~J.}\ \bibnamefont
  {Flatt}}, \ and\ \bibinfo {author} {\bibfnamefont {P.}~\bibnamefont
  {Coussot}},\ }\href@noop {} {\bibfield  {journal} {\bibinfo  {journal}
  {Cement and Concrete Research}\ }\textbf {\bibinfo {volume} {40}},\ \bibinfo
  {pages} {77} (\bibinfo {year} {2010})}\BibitemShut {NoStop}%
\bibitem [{\citenamefont {Brown}\ and\ \citenamefont
  {Jaeger}(2014)}]{Brown_2014}%
  \BibitemOpen
  \bibfield  {author} {\bibinfo {author} {\bibfnamefont {E.}~\bibnamefont
  {Brown}}\ and\ \bibinfo {author} {\bibfnamefont {H.~M.}\ \bibnamefont
  {Jaeger}},\ }\href@noop {} {\bibfield  {journal} {\bibinfo  {journal}
  {Reports on Progress in Physics}\ }\textbf {\bibinfo {volume} {77}},\
  \bibinfo {pages} {046602} (\bibinfo {year} {2014})}\BibitemShut {NoStop}%
\bibitem [{\citenamefont {Peters}\ \emph {et~al.}(2016)\citenamefont {Peters},
  \citenamefont {Majumdar},\ and\ \citenamefont {Jaeger}}]{Peters_2016}%
  \BibitemOpen
  \bibfield  {author} {\bibinfo {author} {\bibfnamefont {I.~R.}\ \bibnamefont
  {Peters}}, \bibinfo {author} {\bibfnamefont {S.}~\bibnamefont {Majumdar}}, \
  and\ \bibinfo {author} {\bibfnamefont {H.~M.}\ \bibnamefont {Jaeger}},\
  }\href@noop {} {\bibfield  {journal} {\bibinfo  {journal} {Nature}\ }\textbf
  {\bibinfo {volume} {532}},\ \bibinfo {pages} {214} (\bibinfo {year}
  {2016})}\BibitemShut {NoStop}%
\bibitem [{\citenamefont {Singh}\ \emph {et~al.}(2019)\citenamefont {Singh},
  \citenamefont {Pednekar}, \citenamefont {Chun}, \citenamefont {Denn},\ and\
  \citenamefont {Morris}}]{Singh_2019}%
  \BibitemOpen
  \bibfield  {author} {\bibinfo {author} {\bibfnamefont {A.}~\bibnamefont
  {Singh}}, \bibinfo {author} {\bibfnamefont {S.}~\bibnamefont {Pednekar}},
  \bibinfo {author} {\bibfnamefont {J.}~\bibnamefont {Chun}}, \bibinfo {author}
  {\bibfnamefont {M.~M.}\ \bibnamefont {Denn}}, \ and\ \bibinfo {author}
  {\bibfnamefont {J.~F.}\ \bibnamefont {Morris}},\ }\href@noop {} {\bibfield
  {journal} {\bibinfo  {journal} {Physical Review Letters}\ }\textbf {\bibinfo
  {volume} {122}},\ \bibinfo {pages} {098004} (\bibinfo {year}
  {2019})}\BibitemShut {NoStop}%
\bibitem [{\citenamefont {Guy}\ \emph {et~al.}(2018)\citenamefont {Guy},
  \citenamefont {Richards}, \citenamefont {Hodgson}, \citenamefont {Blanco},\
  and\ \citenamefont {Poon}}]{Guy_2018}%
  \BibitemOpen
  \bibfield  {author} {\bibinfo {author} {\bibfnamefont {B.~M.}\ \bibnamefont
  {Guy}}, \bibinfo {author} {\bibfnamefont {J.}~\bibnamefont {Richards}},
  \bibinfo {author} {\bibfnamefont {D.}~\bibnamefont {Hodgson}}, \bibinfo
  {author} {\bibfnamefont {E.}~\bibnamefont {Blanco}}, \ and\ \bibinfo {author}
  {\bibfnamefont {W.~C.~K.}\ \bibnamefont {Poon}},\ }\href@noop {} {\bibfield
  {journal} {\bibinfo  {journal} {Physical Review Letters}\ }\textbf {\bibinfo
  {volume} {121}},\ \bibinfo {pages} {128001} (\bibinfo {year}
  {2018})}\BibitemShut {NoStop}%
\bibitem [{\citenamefont {Singh}\ \emph {et~al.}(2020)\citenamefont {Singh},
  \citenamefont {Ness}, \citenamefont {Seto}, \citenamefont {de~Pablo},\ and\
  \citenamefont {Jaeger}}]{Singh_2020}%
  \BibitemOpen
  \bibfield  {author} {\bibinfo {author} {\bibfnamefont {A.}~\bibnamefont
  {Singh}}, \bibinfo {author} {\bibfnamefont {C.}~\bibnamefont {Ness}},
  \bibinfo {author} {\bibfnamefont {R.}~\bibnamefont {Seto}}, \bibinfo {author}
  {\bibfnamefont {J.~J.}\ \bibnamefont {de~Pablo}}, \ and\ \bibinfo {author}
  {\bibfnamefont {H.~M.}\ \bibnamefont {Jaeger}},\ }\href@noop {} {\bibfield
  {journal} {\bibinfo  {journal} {Physical Review Letters}\ }\textbf {\bibinfo
  {volume} {124}},\ \bibinfo {pages} {248005} (\bibinfo {year}
  {2020})}\BibitemShut {NoStop}%
\bibitem [{\citenamefont {Singh}\ \emph {et~al.}(2022)\citenamefont {Singh},
  \citenamefont {Jackson}, \citenamefont {van~der Naald}, \citenamefont
  {de~Pablo},\ and\ \citenamefont {Jaeger}}]{Singh_2022}%
  \BibitemOpen
  \bibfield  {author} {\bibinfo {author} {\bibfnamefont {A.}~\bibnamefont
  {Singh}}, \bibinfo {author} {\bibfnamefont {G.~L.}\ \bibnamefont {Jackson}},
  \bibinfo {author} {\bibfnamefont {M.}~\bibnamefont {van~der Naald}}, \bibinfo
  {author} {\bibfnamefont {J.~J.}\ \bibnamefont {de~Pablo}}, \ and\ \bibinfo
  {author} {\bibfnamefont {H.~M.}\ \bibnamefont {Jaeger}},\ }\href@noop {}
  {\bibfield  {journal} {\bibinfo  {journal} {Physical Review Fluids}\ }\textbf
  {\bibinfo {volume} {7}},\ \bibinfo {pages} {054302} (\bibinfo {year}
  {2022})}\BibitemShut {NoStop}%
\bibitem [{\citenamefont {Seto}\ \emph {et~al.}(2013)\citenamefont {Seto},
  \citenamefont {Mari}, \citenamefont {Morris},\ and\ \citenamefont
  {Denn}}]{Seto_2013a}%
  \BibitemOpen
  \bibfield  {author} {\bibinfo {author} {\bibfnamefont {R.}~\bibnamefont
  {Seto}}, \bibinfo {author} {\bibfnamefont {R.}~\bibnamefont {Mari}}, \bibinfo
  {author} {\bibfnamefont {J.~F.}\ \bibnamefont {Morris}}, \ and\ \bibinfo
  {author} {\bibfnamefont {M.~M.}\ \bibnamefont {Denn}},\ }\href@noop {}
  {\bibfield  {journal} {\bibinfo  {journal} {Physical Review Letters}\
  }\textbf {\bibinfo {volume} {111}},\ \bibinfo {pages} {218301} (\bibinfo
  {year} {2013})}\BibitemShut {NoStop}%
\bibitem [{\citenamefont {Wyart}\ and\ \citenamefont
  {Cates}(2014)}]{Wyart_2014}%
  \BibitemOpen
  \bibfield  {author} {\bibinfo {author} {\bibfnamefont {M.}~\bibnamefont
  {Wyart}}\ and\ \bibinfo {author} {\bibfnamefont {M.~E.}\ \bibnamefont
  {Cates}},\ }\href@noop {} {\bibfield  {journal} {\bibinfo  {journal}
  {Physical Review Letters}\ }\textbf {\bibinfo {volume} {112}},\ \bibinfo
  {pages} {098302} (\bibinfo {year} {2014})}\BibitemShut {NoStop}%
\bibitem [{\citenamefont {Singh}\ \emph {et~al.}(2018)\citenamefont {Singh},
  \citenamefont {Mari}, \citenamefont {Denn},\ and\ \citenamefont
  {Morris}}]{Singh_2018}%
  \BibitemOpen
  \bibfield  {author} {\bibinfo {author} {\bibfnamefont {A.}~\bibnamefont
  {Singh}}, \bibinfo {author} {\bibfnamefont {R.}~\bibnamefont {Mari}},
  \bibinfo {author} {\bibfnamefont {M.~M.}\ \bibnamefont {Denn}}, \ and\
  \bibinfo {author} {\bibfnamefont {J.~F.}\ \bibnamefont {Morris}},\
  }\href@noop {} {\bibfield  {journal} {\bibinfo  {journal} {Journal of
  Rheology}\ }\textbf {\bibinfo {volume} {62}},\ \bibinfo {pages} {457}
  (\bibinfo {year} {2018})}\BibitemShut {NoStop}%
\bibitem [{\citenamefont {Guy}\ \emph {et~al.}(2015)\citenamefont {Guy},
  \citenamefont {Hermes},\ and\ \citenamefont {Poon}}]{Guy_2015}%
  \BibitemOpen
  \bibfield  {author} {\bibinfo {author} {\bibfnamefont {B.~M.}\ \bibnamefont
  {Guy}}, \bibinfo {author} {\bibfnamefont {M.}~\bibnamefont {Hermes}}, \ and\
  \bibinfo {author} {\bibfnamefont {W.~C.~K.}\ \bibnamefont {Poon}},\
  }\href@noop {} {\bibfield  {journal} {\bibinfo  {journal} {Physical Review
  Letters}\ }\textbf {\bibinfo {volume} {115}},\ \bibinfo {pages} {088304}
  (\bibinfo {year} {2015})}\BibitemShut {NoStop}%
\bibitem [{\citenamefont {Ness}\ and\ \citenamefont {Sun}(2016)}]{Ness_2016}%
  \BibitemOpen
  \bibfield  {author} {\bibinfo {author} {\bibfnamefont {C.}~\bibnamefont
  {Ness}}\ and\ \bibinfo {author} {\bibfnamefont {J.}~\bibnamefont {Sun}},\
  }\href@noop {} {\bibfield  {journal} {\bibinfo  {journal} {Soft Matter}\
  }\textbf {\bibinfo {volume} {12}},\ \bibinfo {pages} {914} (\bibinfo {year}
  {2016})}\BibitemShut {NoStop}%
\bibitem [{\citenamefont {Guy}\ \emph {et~al.}(2020)\citenamefont {Guy},
  \citenamefont {Ness}, \citenamefont {Hermes}, \citenamefont {Sawiak},
  \citenamefont {Sun},\ and\ \citenamefont {Poon}}]{Guy_2020}%
  \BibitemOpen
  \bibfield  {author} {\bibinfo {author} {\bibfnamefont {B.~M.}\ \bibnamefont
  {Guy}}, \bibinfo {author} {\bibfnamefont {C.}~\bibnamefont {Ness}}, \bibinfo
  {author} {\bibfnamefont {M.}~\bibnamefont {Hermes}}, \bibinfo {author}
  {\bibfnamefont {L.~J.}\ \bibnamefont {Sawiak}}, \bibinfo {author}
  {\bibfnamefont {J.}~\bibnamefont {Sun}}, \ and\ \bibinfo {author}
  {\bibfnamefont {W.~C.}\ \bibnamefont {Poon}},\ }\href@noop {} {\bibfield
  {journal} {\bibinfo  {journal} {Soft Matter}\ }\textbf {\bibinfo {volume}
  {16}},\ \bibinfo {pages} {229} (\bibinfo {year} {2020})}\BibitemShut
  {NoStop}%
\bibitem [{\citenamefont {Pradeep}\ \emph {et~al.}(2021)\citenamefont
  {Pradeep}, \citenamefont {Nabizadeh}, \citenamefont {Jacob}, \citenamefont
  {Jamali},\ and\ \citenamefont {Hsiao}}]{Pradeep_2021}%
  \BibitemOpen
  \bibfield  {author} {\bibinfo {author} {\bibfnamefont {S.}~\bibnamefont
  {Pradeep}}, \bibinfo {author} {\bibfnamefont {M.}~\bibnamefont {Nabizadeh}},
  \bibinfo {author} {\bibfnamefont {A.~R.}\ \bibnamefont {Jacob}}, \bibinfo
  {author} {\bibfnamefont {S.}~\bibnamefont {Jamali}}, \ and\ \bibinfo {author}
  {\bibfnamefont {L.~C.}\ \bibnamefont {Hsiao}},\ }\href@noop {} {\bibfield
  {journal} {\bibinfo  {journal} {Physical Review Letters}\ }\textbf {\bibinfo
  {volume} {127}},\ \bibinfo {pages} {158002} (\bibinfo {year}
  {2021})}\BibitemShut {NoStop}%
\bibitem [{\citenamefont {Van~Damme}(2018)}]{van2018concrete}%
  \BibitemOpen
  \bibfield  {author} {\bibinfo {author} {\bibfnamefont {H.}~\bibnamefont
  {Van~Damme}},\ }\href@noop {} {\bibfield  {journal} {\bibinfo  {journal}
  {Cement and Concrete Research}\ }\textbf {\bibinfo {volume} {112}},\ \bibinfo
  {pages} {5} (\bibinfo {year} {2018})}\BibitemShut {NoStop}%
\bibitem [{\citenamefont {Flatt}(2004)}]{flatt2004towards}%
  \BibitemOpen
  \bibfield  {author} {\bibinfo {author} {\bibfnamefont {R.}~\bibnamefont
  {Flatt}},\ }\href@noop {} {\bibfield  {journal} {\bibinfo  {journal}
  {Materials and Structures}\ }\textbf {\bibinfo {volume} {37}},\ \bibinfo
  {pages} {289} (\bibinfo {year} {2004})}\BibitemShut {NoStop}%
\bibitem [{\citenamefont {Cwalina}\ and\ \citenamefont
  {Wagner}(2016)}]{cwalina2016rheology}%
  \BibitemOpen
  \bibfield  {author} {\bibinfo {author} {\bibfnamefont {C.~D.}\ \bibnamefont
  {Cwalina}}\ and\ \bibinfo {author} {\bibfnamefont {N.~J.}\ \bibnamefont
  {Wagner}},\ }\href@noop {} {\bibfield  {journal} {\bibinfo  {journal}
  {Journal of Rheology}\ }\textbf {\bibinfo {volume} {60}},\ \bibinfo {pages}
  {47} (\bibinfo {year} {2016})}\BibitemShut {NoStop}%
\bibitem [{\citenamefont {Madraki}\ \emph {et~al.}(2017)\citenamefont
  {Madraki}, \citenamefont {Hormozi}, \citenamefont {Ovarlez}, \citenamefont
  {Guazzelli},\ and\ \citenamefont {Pouliquen}}]{Madraki_2017}%
  \BibitemOpen
  \bibfield  {author} {\bibinfo {author} {\bibfnamefont {Y.}~\bibnamefont
  {Madraki}}, \bibinfo {author} {\bibfnamefont {S.}~\bibnamefont {Hormozi}},
  \bibinfo {author} {\bibfnamefont {G.}~\bibnamefont {Ovarlez}}, \bibinfo
  {author} {\bibfnamefont {E.}~\bibnamefont {Guazzelli}}, \ and\ \bibinfo
  {author} {\bibfnamefont {O.}~\bibnamefont {Pouliquen}},\ }\href@noop {}
  {\bibfield  {journal} {\bibinfo  {journal} {Physical Review Fluids}\ }\textbf
  {\bibinfo {volume} {2}},\ \bibinfo {pages} {033301} (\bibinfo {year}
  {2017})}\BibitemShut {NoStop}%
\bibitem [{\citenamefont {Madraki}\ \emph {et~al.}(2018)\citenamefont
  {Madraki}, \citenamefont {Ovarlez},\ and\ \citenamefont
  {Hormozi}}]{Madraki_2018}%
  \BibitemOpen
  \bibfield  {author} {\bibinfo {author} {\bibfnamefont {Y.}~\bibnamefont
  {Madraki}}, \bibinfo {author} {\bibfnamefont {G.}~\bibnamefont {Ovarlez}}, \
  and\ \bibinfo {author} {\bibfnamefont {S.}~\bibnamefont {Hormozi}},\
  }\href@noop {} {\bibfield  {journal} {\bibinfo  {journal} {Physical Review
  Letters}\ }\textbf {\bibinfo {volume} {121}},\ \bibinfo {pages} {108001}
  (\bibinfo {year} {2018})}\BibitemShut {NoStop}%
\bibitem [{\citenamefont {Malbranche}\ \emph {et~al.}(2023)\citenamefont
  {Malbranche}, \citenamefont {Chakraborty},\ and\ \citenamefont
  {Morris}}]{Malbranche_2023}%
  \BibitemOpen
  \bibfield  {author} {\bibinfo {author} {\bibfnamefont {N.}~\bibnamefont
  {Malbranche}}, \bibinfo {author} {\bibfnamefont {B.}~\bibnamefont
  {Chakraborty}}, \ and\ \bibinfo {author} {\bibfnamefont {J.~F.}\ \bibnamefont
  {Morris}},\ }\href@noop {} {\bibfield  {journal} {\bibinfo  {journal}
  {Journal of Rheology}\ }\textbf {\bibinfo {volume} {67}},\ \bibinfo {pages}
  {91} (\bibinfo {year} {2023})}\BibitemShut {NoStop}%
\bibitem [{\citenamefont {Pednekar}\ \emph {et~al.}(2018)\citenamefont
  {Pednekar}, \citenamefont {Chun},\ and\ \citenamefont
  {Morris}}]{Pednekar_2018}%
  \BibitemOpen
  \bibfield  {author} {\bibinfo {author} {\bibfnamefont {S.}~\bibnamefont
  {Pednekar}}, \bibinfo {author} {\bibfnamefont {J.}~\bibnamefont {Chun}}, \
  and\ \bibinfo {author} {\bibfnamefont {J.~F.}\ \bibnamefont {Morris}},\
  }\href@noop {} {\bibfield  {journal} {\bibinfo  {journal} {Journal of
  Rheology}\ }\textbf {\bibinfo {volume} {62}},\ \bibinfo {pages} {513}
  (\bibinfo {year} {2018})}\BibitemShut {NoStop}%
\bibitem [{\citenamefont {Monti}\ and\ \citenamefont
  {Rosti}(2023)}]{monti2023shear}%
  \BibitemOpen
  \bibfield  {author} {\bibinfo {author} {\bibfnamefont {A.}~\bibnamefont
  {Monti}}\ and\ \bibinfo {author} {\bibfnamefont {M.~E.}\ \bibnamefont
  {Rosti}},\ }\href@noop {} {\bibfield  {journal} {\bibinfo  {journal}
  {Meccanica}\ }\textbf {\bibinfo {volume} {58}},\ \bibinfo {pages} {727}
  (\bibinfo {year} {2023})}\BibitemShut {NoStop}%
\bibitem [{\citenamefont {Malbranche}\ \emph {et~al.}(2022)\citenamefont
  {Malbranche}, \citenamefont {Santra}, \citenamefont {Chakraborty},\ and\
  \citenamefont {Morris}}]{malbranche2022scaling}%
  \BibitemOpen
  \bibfield  {author} {\bibinfo {author} {\bibfnamefont {N.}~\bibnamefont
  {Malbranche}}, \bibinfo {author} {\bibfnamefont {A.}~\bibnamefont {Santra}},
  \bibinfo {author} {\bibfnamefont {B.}~\bibnamefont {Chakraborty}}, \ and\
  \bibinfo {author} {\bibfnamefont {J.~F.}\ \bibnamefont {Morris}},\
  }\href@noop {} {\bibfield  {journal} {\bibinfo  {journal} {Frontiers in
  physics}\ }\textbf {\bibinfo {volume} {10}},\ \bibinfo {pages} {946221}
  (\bibinfo {year} {2022})}\BibitemShut {NoStop}%
\bibitem [{\citenamefont {Ramaswamy}\ \emph {et~al.}(2023)\citenamefont
  {Ramaswamy}, \citenamefont {Griniasty}, \citenamefont {Liarte}, \citenamefont
  {Shetty}, \citenamefont {Katifori}, \citenamefont {Del~Gado}, \citenamefont
  {Sethna}, \citenamefont {Chakraborty},\ and\ \citenamefont
  {Cohen}}]{Ramaswamy_2023}%
  \BibitemOpen
  \bibfield  {author} {\bibinfo {author} {\bibfnamefont {M.}~\bibnamefont
  {Ramaswamy}}, \bibinfo {author} {\bibfnamefont {I.}~\bibnamefont
  {Griniasty}}, \bibinfo {author} {\bibfnamefont {D.~B.}\ \bibnamefont
  {Liarte}}, \bibinfo {author} {\bibfnamefont {A.}~\bibnamefont {Shetty}},
  \bibinfo {author} {\bibfnamefont {E.}~\bibnamefont {Katifori}}, \bibinfo
  {author} {\bibfnamefont {E.}~\bibnamefont {Del~Gado}}, \bibinfo {author}
  {\bibfnamefont {J.~P.}\ \bibnamefont {Sethna}}, \bibinfo {author}
  {\bibfnamefont {B.}~\bibnamefont {Chakraborty}}, \ and\ \bibinfo {author}
  {\bibfnamefont {I.}~\bibnamefont {Cohen}},\ }\href@noop {} {\bibfield
  {journal} {\bibinfo  {journal} {Journal of Rheology}\ }\textbf {\bibinfo
  {volume} {67}},\ \bibinfo {pages} {1189} (\bibinfo {year}
  {2023})}\BibitemShut {NoStop}%
\bibitem [{\citenamefont {Maranzano}\ and\ \citenamefont
  {Wagner}(2001{\natexlab{a}})}]{Maranzano_2001}%
  \BibitemOpen
  \bibfield  {author} {\bibinfo {author} {\bibfnamefont {B.~J.}\ \bibnamefont
  {Maranzano}}\ and\ \bibinfo {author} {\bibfnamefont {N.~J.}\ \bibnamefont
  {Wagner}},\ }\href@noop {} {\bibfield  {journal} {\bibinfo  {journal}
  {Journal of Chemical Physics}\ }\textbf {\bibinfo {volume} {114}},\ \bibinfo
  {pages} {10514} (\bibinfo {year} {2001}{\natexlab{a}})}\BibitemShut {NoStop}%
\bibitem [{\citenamefont {Farris}(1968)}]{Farris_1968}%
  \BibitemOpen
  \bibfield  {author} {\bibinfo {author} {\bibfnamefont {R.}~\bibnamefont
  {Farris}},\ }\href@noop {} {\bibfield  {journal} {\bibinfo  {journal}
  {Transactions of the Society of Rheology}\ }\textbf {\bibinfo {volume}
  {12}},\ \bibinfo {pages} {281} (\bibinfo {year} {1968})}\BibitemShut
  {NoStop}%
\bibitem [{\citenamefont {Mwasame}\ \emph {et~al.}(2016)\citenamefont
  {Mwasame}, \citenamefont {Wagner},\ and\ \citenamefont
  {Beris}}]{mwasame2016modeling}%
  \BibitemOpen
  \bibfield  {author} {\bibinfo {author} {\bibfnamefont {P.~M.}\ \bibnamefont
  {Mwasame}}, \bibinfo {author} {\bibfnamefont {N.~J.}\ \bibnamefont {Wagner}},
  \ and\ \bibinfo {author} {\bibfnamefont {A.~N.}\ \bibnamefont {Beris}},\
  }\href@noop {} {\bibfield  {journal} {\bibinfo  {journal} {Journal of
  Rheology}\ }\textbf {\bibinfo {volume} {60}},\ \bibinfo {pages} {225}
  (\bibinfo {year} {2016})}\BibitemShut {NoStop}%
\bibitem [{\citenamefont {Shapiro}\ and\ \citenamefont
  {Probstein}(1992)}]{Shapiro_1992}%
  \BibitemOpen
  \bibfield  {author} {\bibinfo {author} {\bibfnamefont {A.~P.}\ \bibnamefont
  {Shapiro}}\ and\ \bibinfo {author} {\bibfnamefont {R.~F.}\ \bibnamefont
  {Probstein}},\ }\href@noop {} {\bibfield  {journal} {\bibinfo  {journal}
  {Physical review letters}\ }\textbf {\bibinfo {volume} {68}},\ \bibinfo
  {pages} {1422} (\bibinfo {year} {1992})}\BibitemShut {NoStop}%
\bibitem [{\citenamefont {Chong}\ \emph {et~al.}(1971)\citenamefont {Chong},
  \citenamefont {Christiansen},\ and\ \citenamefont {Baer}}]{Chong_1971}%
  \BibitemOpen
  \bibfield  {author} {\bibinfo {author} {\bibfnamefont {J.}~\bibnamefont
  {Chong}}, \bibinfo {author} {\bibfnamefont {E.}~\bibnamefont {Christiansen}},
  \ and\ \bibinfo {author} {\bibfnamefont {A.}~\bibnamefont {Baer}},\
  }\href@noop {} {\bibfield  {journal} {\bibinfo  {journal} {Journal of applied
  polymer science}\ }\textbf {\bibinfo {volume} {15}},\ \bibinfo {pages} {2007}
  (\bibinfo {year} {1971})}\BibitemShut {NoStop}%
\bibitem [{\citenamefont {Barnes}(1989)}]{Barnes_1989}%
  \BibitemOpen
  \bibfield  {author} {\bibinfo {author} {\bibfnamefont {H.~A.}\ \bibnamefont
  {Barnes}},\ }\href@noop {} {\bibfield  {journal} {\bibinfo  {journal} {J.
  Rheol.}\ }\textbf {\bibinfo {volume} {33}},\ \bibinfo {pages} {329} (\bibinfo
  {year} {1989})}\BibitemShut {NoStop}%
\bibitem [{\citenamefont {Gondret}\ and\ \citenamefont
  {Petit}(1997)}]{Gondret_1997}%
  \BibitemOpen
  \bibfield  {author} {\bibinfo {author} {\bibfnamefont {P.}~\bibnamefont
  {Gondret}}\ and\ \bibinfo {author} {\bibfnamefont {L.}~\bibnamefont
  {Petit}},\ }\href@noop {} {\bibfield  {journal} {\bibinfo  {journal} {Journal
  of rheology}\ }\textbf {\bibinfo {volume} {41}},\ \bibinfo {pages} {1261}
  (\bibinfo {year} {1997})}\BibitemShut {NoStop}%
\bibitem [{\citenamefont {Poslinski}\ \emph {et~al.}(1988)\citenamefont
  {Poslinski}, \citenamefont {Ryan}, \citenamefont {Gupta}, \citenamefont
  {Seshadri},\ and\ \citenamefont {Frechette}}]{Poslinski_1988}%
  \BibitemOpen
  \bibfield  {author} {\bibinfo {author} {\bibfnamefont {A.}~\bibnamefont
  {Poslinski}}, \bibinfo {author} {\bibfnamefont {M.}~\bibnamefont {Ryan}},
  \bibinfo {author} {\bibfnamefont {R.}~\bibnamefont {Gupta}}, \bibinfo
  {author} {\bibfnamefont {S.}~\bibnamefont {Seshadri}}, \ and\ \bibinfo
  {author} {\bibfnamefont {F.}~\bibnamefont {Frechette}},\ }\href@noop {}
  {\bibfield  {journal} {\bibinfo  {journal} {Journal of Rheology}\ }\textbf
  {\bibinfo {volume} {32}},\ \bibinfo {pages} {703} (\bibinfo {year}
  {1988})}\BibitemShut {NoStop}%
\bibitem [{\citenamefont {Maranzano}\ and\ \citenamefont
  {Wagner}(2001{\natexlab{b}})}]{Maranzano_2001a}%
  \BibitemOpen
  \bibfield  {author} {\bibinfo {author} {\bibfnamefont {B.~J.}\ \bibnamefont
  {Maranzano}}\ and\ \bibinfo {author} {\bibfnamefont {N.~J.}\ \bibnamefont
  {Wagner}},\ }\href@noop {} {\bibfield  {journal} {\bibinfo  {journal}
  {Journal of Rheology}\ }\textbf {\bibinfo {volume} {45}},\ \bibinfo {pages}
  {1205} (\bibinfo {year} {2001}{\natexlab{b}})}\BibitemShut {NoStop}%
\bibitem [{\citenamefont {Chang}\ and\ \citenamefont
  {Powell}(1994{\natexlab{a}})}]{Chang_1994}%
  \BibitemOpen
  \bibfield  {author} {\bibinfo {author} {\bibfnamefont {C.}~\bibnamefont
  {Chang}}\ and\ \bibinfo {author} {\bibfnamefont {R.~L.}\ \bibnamefont
  {Powell}},\ }\href@noop {} {\bibfield  {journal} {\bibinfo  {journal}
  {Journal of rheology}\ }\textbf {\bibinfo {volume} {38}},\ \bibinfo {pages}
  {85} (\bibinfo {year} {1994}{\natexlab{a}})}\BibitemShut {NoStop}%
\bibitem [{\citenamefont {Chang}\ and\ \citenamefont
  {Powell}(1994{\natexlab{b}})}]{Chang_1994a}%
  \BibitemOpen
  \bibfield  {author} {\bibinfo {author} {\bibfnamefont {C.}~\bibnamefont
  {Chang}}\ and\ \bibinfo {author} {\bibfnamefont {R.~L.}\ \bibnamefont
  {Powell}},\ }\href@noop {} {\bibfield  {journal} {\bibinfo  {journal}
  {Physics of Fluids}\ }\textbf {\bibinfo {volume} {6}},\ \bibinfo {pages}
  {1628} (\bibinfo {year} {1994}{\natexlab{b}})}\BibitemShut {NoStop}%
\bibitem [{\citenamefont {Chang}\ and\ \citenamefont
  {Powell}(1993)}]{Chang_1993}%
  \BibitemOpen
  \bibfield  {author} {\bibinfo {author} {\bibfnamefont {C.}~\bibnamefont
  {Chang}}\ and\ \bibinfo {author} {\bibfnamefont {R.~L.}\ \bibnamefont
  {Powell}},\ }\href@noop {} {\bibfield  {journal} {\bibinfo  {journal}
  {Journal of Fluid Mechanics}\ }\textbf {\bibinfo {volume} {253}},\ \bibinfo
  {pages} {1} (\bibinfo {year} {1993})}\BibitemShut {NoStop}%
\bibitem [{\citenamefont {Yu}\ and\ \citenamefont
  {Standish}(1991)}]{yu1991estimation}%
  \BibitemOpen
  \bibfield  {author} {\bibinfo {author} {\bibfnamefont {A.-B.}\ \bibnamefont
  {Yu}}\ and\ \bibinfo {author} {\bibfnamefont {N.}~\bibnamefont {Standish}},\
  }\href@noop {} {\bibfield  {journal} {\bibinfo  {journal} {Industrial \&
  Engineering Chemistry Research}\ }\textbf {\bibinfo {volume} {30}},\ \bibinfo
  {pages} {1372} (\bibinfo {year} {1991})}\BibitemShut {NoStop}%
\bibitem [{\citenamefont {Qi}\ and\ \citenamefont
  {Tanner}(2011)}]{qi2011relative}%
  \BibitemOpen
  \bibfield  {author} {\bibinfo {author} {\bibfnamefont {F.}~\bibnamefont
  {Qi}}\ and\ \bibinfo {author} {\bibfnamefont {R.~I.}\ \bibnamefont
  {Tanner}},\ }\href@noop {} {\bibfield  {journal} {\bibinfo  {journal}
  {Korea-Australia Rheology Journal}\ }\textbf {\bibinfo {volume} {23}},\
  \bibinfo {pages} {105} (\bibinfo {year} {2011})}\BibitemShut {NoStop}%
\bibitem [{\citenamefont {Anzivino}\ \emph {et~al.}(2023)\citenamefont
  {Anzivino}, \citenamefont {Casiulis}, \citenamefont {Zhang}, \citenamefont
  {Moussa}, \citenamefont {Martiniani},\ and\ \citenamefont
  {Zaccone}}]{anzivino2023estimating}%
  \BibitemOpen
  \bibfield  {author} {\bibinfo {author} {\bibfnamefont {C.}~\bibnamefont
  {Anzivino}}, \bibinfo {author} {\bibfnamefont {M.}~\bibnamefont {Casiulis}},
  \bibinfo {author} {\bibfnamefont {T.}~\bibnamefont {Zhang}}, \bibinfo
  {author} {\bibfnamefont {A.~S.}\ \bibnamefont {Moussa}}, \bibinfo {author}
  {\bibfnamefont {S.}~\bibnamefont {Martiniani}}, \ and\ \bibinfo {author}
  {\bibfnamefont {A.}~\bibnamefont {Zaccone}},\ }\href@noop {} {\bibfield
  {journal} {\bibinfo  {journal} {The Journal of Chemical Physics}\ }\textbf
  {\bibinfo {volume} {158}} (\bibinfo {year} {2023})}\BibitemShut {NoStop}%
\bibitem [{\citenamefont {Servais}\ \emph {et~al.}(2002)\citenamefont
  {Servais}, \citenamefont {Jones},\ and\ \citenamefont
  {Roberts}}]{servais2002influence}%
  \BibitemOpen
  \bibfield  {author} {\bibinfo {author} {\bibfnamefont {C.}~\bibnamefont
  {Servais}}, \bibinfo {author} {\bibfnamefont {R.}~\bibnamefont {Jones}}, \
  and\ \bibinfo {author} {\bibfnamefont {I.}~\bibnamefont {Roberts}},\
  }\href@noop {} {\bibfield  {journal} {\bibinfo  {journal} {Journal of food
  engineering}\ }\textbf {\bibinfo {volume} {51}},\ \bibinfo {pages} {201}
  (\bibinfo {year} {2002})}\BibitemShut {NoStop}%
\bibitem [{\citenamefont {Pishvaei}\ \emph {et~al.}(2006)\citenamefont
  {Pishvaei}, \citenamefont {Graillat}, \citenamefont {Cassagnau},\ and\
  \citenamefont {McKenna}}]{pishvaei2006modelling}%
  \BibitemOpen
  \bibfield  {author} {\bibinfo {author} {\bibfnamefont {M.}~\bibnamefont
  {Pishvaei}}, \bibinfo {author} {\bibfnamefont {C.}~\bibnamefont {Graillat}},
  \bibinfo {author} {\bibfnamefont {P.}~\bibnamefont {Cassagnau}}, \ and\
  \bibinfo {author} {\bibfnamefont {T.}~\bibnamefont {McKenna}},\ }\href@noop
  {} {\bibfield  {journal} {\bibinfo  {journal} {Chemical engineering science}\
  }\textbf {\bibinfo {volume} {61}},\ \bibinfo {pages} {5768} (\bibinfo {year}
  {2006})}\BibitemShut {NoStop}%
\bibitem [{\citenamefont {Srivastava}\ \emph {et~al.}(2021)\citenamefont
  {Srivastava}, \citenamefont {Roberts}, \citenamefont {Clemmer}, \citenamefont
  {Silbert}, \citenamefont {Lechman},\ and\ \citenamefont
  {Grest}}]{Srivastava_2021}%
  \BibitemOpen
  \bibfield  {author} {\bibinfo {author} {\bibfnamefont {I.}~\bibnamefont
  {Srivastava}}, \bibinfo {author} {\bibfnamefont {S.~A.}\ \bibnamefont
  {Roberts}}, \bibinfo {author} {\bibfnamefont {J.~T.}\ \bibnamefont
  {Clemmer}}, \bibinfo {author} {\bibfnamefont {L.~E.}\ \bibnamefont
  {Silbert}}, \bibinfo {author} {\bibfnamefont {J.~B.}\ \bibnamefont
  {Lechman}}, \ and\ \bibinfo {author} {\bibfnamefont {G.~S.}\ \bibnamefont
  {Grest}},\ }\href@noop {} {\bibfield  {journal} {\bibinfo  {journal}
  {Physical Review Research}\ }\textbf {\bibinfo {volume} {3}},\ \bibinfo
  {pages} {L032042} (\bibinfo {year} {2021})}\BibitemShut {NoStop}%
\bibitem [{\citenamefont {Petit}\ \emph {et~al.}(2020)\citenamefont {Petit},
  \citenamefont {Kumar}, \citenamefont {Luding},\ and\ \citenamefont
  {Sperl}}]{petit2020additional}%
  \BibitemOpen
  \bibfield  {author} {\bibinfo {author} {\bibfnamefont {J.~C.}\ \bibnamefont
  {Petit}}, \bibinfo {author} {\bibfnamefont {N.}~\bibnamefont {Kumar}},
  \bibinfo {author} {\bibfnamefont {S.}~\bibnamefont {Luding}}, \ and\ \bibinfo
  {author} {\bibfnamefont {M.}~\bibnamefont {Sperl}},\ }\href@noop {}
  {\bibfield  {journal} {\bibinfo  {journal} {Physical Review Letters}\
  }\textbf {\bibinfo {volume} {125}},\ \bibinfo {pages} {215501} (\bibinfo
  {year} {2020})}\BibitemShut {NoStop}%
\bibitem [{\citenamefont {Mari}\ \emph {et~al.}(2015)\citenamefont {Mari},
  \citenamefont {Seto}, \citenamefont {Morris},\ and\ \citenamefont
  {Denn}}]{Mari_2015}%
  \BibitemOpen
  \bibfield  {author} {\bibinfo {author} {\bibfnamefont {R.}~\bibnamefont
  {Mari}}, \bibinfo {author} {\bibfnamefont {R.}~\bibnamefont {Seto}}, \bibinfo
  {author} {\bibfnamefont {J.~F.}\ \bibnamefont {Morris}}, \ and\ \bibinfo
  {author} {\bibfnamefont {M.~M.}\ \bibnamefont {Denn}},\ }\href@noop {}
  {\bibfield  {journal} {\bibinfo  {journal} {Physical Review E}\ }\textbf
  {\bibinfo {volume} {91}},\ \bibinfo {pages} {052302} (\bibinfo {year}
  {2015})}\BibitemShut {NoStop}%
\bibitem [{\citenamefont {Brady}\ and\ \citenamefont
  {Bossis}(1985)}]{Brady_1985}%
  \BibitemOpen
  \bibfield  {author} {\bibinfo {author} {\bibfnamefont {J.~F.}\ \bibnamefont
  {Brady}}\ and\ \bibinfo {author} {\bibfnamefont {G.}~\bibnamefont {Bossis}},\
  }\href@noop {} {\bibfield  {journal} {\bibinfo  {journal} {J. Fluid Mech.}\
  }\textbf {\bibinfo {volume} {155}},\ \bibinfo {pages} {105} (\bibinfo {year}
  {1985})}\BibitemShut {NoStop}%
\bibitem [{\citenamefont {Ball}\ and\ \citenamefont
  {Melrose}(1997)}]{Ball_1997}%
  \BibitemOpen
  \bibfield  {author} {\bibinfo {author} {\bibfnamefont {R.~C.}\ \bibnamefont
  {Ball}}\ and\ \bibinfo {author} {\bibfnamefont {J.~R.}\ \bibnamefont
  {Melrose}},\ }\href@noop {} {\bibfield  {journal} {\bibinfo  {journal} {Phys.
  A}\ }\textbf {\bibinfo {volume} {247}},\ \bibinfo {pages} {444} (\bibinfo
  {year} {1997})}\BibitemShut {NoStop}%
\bibitem [{\citenamefont {Cundall}\ and\ \citenamefont
  {Strack}(1979)}]{Cundall_1979}%
  \BibitemOpen
  \bibfield  {author} {\bibinfo {author} {\bibfnamefont {P.~A.}\ \bibnamefont
  {Cundall}}\ and\ \bibinfo {author} {\bibfnamefont {O.~D.~L.}\ \bibnamefont
  {Strack}},\ }\href@noop {} {\bibfield  {journal} {\bibinfo  {journal}
  {Geotechnique}\ }\textbf {\bibinfo {volume} {29}},\ \bibinfo {pages} {47}
  (\bibinfo {year} {1979})}\BibitemShut {NoStop}%
\bibitem [{\citenamefont {Luding}(2008)}]{Luding_2008}%
  \BibitemOpen
  \bibfield  {author} {\bibinfo {author} {\bibfnamefont {S.}~\bibnamefont
  {Luding}},\ }\href@noop {} {\bibfield  {journal} {\bibinfo  {journal} {Gran.
  Matt.}\ }\textbf {\bibinfo {volume} {10}},\ \bibinfo {pages} {235} (\bibinfo
  {year} {2008})}\BibitemShut {NoStop}%
\bibitem [{\citenamefont {Mari}\ \emph {et~al.}(2014)\citenamefont {Mari},
  \citenamefont {Seto}, \citenamefont {Morris},\ and\ \citenamefont
  {Denn}}]{Mari_2014}%
  \BibitemOpen
  \bibfield  {author} {\bibinfo {author} {\bibfnamefont {R.}~\bibnamefont
  {Mari}}, \bibinfo {author} {\bibfnamefont {R.}~\bibnamefont {Seto}}, \bibinfo
  {author} {\bibfnamefont {J.~F.}\ \bibnamefont {Morris}}, \ and\ \bibinfo
  {author} {\bibfnamefont {M.~M.}\ \bibnamefont {Denn}},\ }\href@noop {}
  {\bibfield  {journal} {\bibinfo  {journal} {Journal of Rheology}\ }\textbf
  {\bibinfo {volume} {58}},\ \bibinfo {pages} {1693} (\bibinfo {year}
  {2014})}\BibitemShut {NoStop}%
\bibitem [{\citenamefont {Krieger}\ and\ \citenamefont
  {Dougherty}(1959)}]{krieger1959mechanism}%
  \BibitemOpen
  \bibfield  {author} {\bibinfo {author} {\bibfnamefont {I.~M.}\ \bibnamefont
  {Krieger}}\ and\ \bibinfo {author} {\bibfnamefont {T.~J.}\ \bibnamefont
  {Dougherty}},\ }\href@noop {} {\bibfield  {journal} {\bibinfo  {journal}
  {Transactions of the Society of Rheology}\ }\textbf {\bibinfo {volume} {3}},\
  \bibinfo {pages} {137} (\bibinfo {year} {1959})}\BibitemShut {NoStop}%
\bibitem [{\citenamefont {Liu}\ and\ \citenamefont
  {Nagel}(2010)}]{liu2010jamming}%
  \BibitemOpen
  \bibfield  {author} {\bibinfo {author} {\bibfnamefont {A.~J.}\ \bibnamefont
  {Liu}}\ and\ \bibinfo {author} {\bibfnamefont {S.~R.}\ \bibnamefont
  {Nagel}},\ }\href@noop {} {\bibfield  {journal} {\bibinfo  {journal} {Annual
  Review of Condensed Matter Physics}\ }\textbf {\bibinfo {volume} {1}},\
  \bibinfo {pages} {347} (\bibinfo {year} {2010})}\BibitemShut {NoStop}%
\bibitem [{\citenamefont {Li}\ \emph {et~al.}(2024)\citenamefont {Li},
  \citenamefont {Royer},\ and\ \citenamefont {Ness}}]{li2023simulating}%
  \BibitemOpen
  \bibfield  {author} {\bibinfo {author} {\bibfnamefont {X.}~\bibnamefont
  {Li}}, \bibinfo {author} {\bibfnamefont {J.~R.}\ \bibnamefont {Royer}}, \
  and\ \bibinfo {author} {\bibfnamefont {C.}~\bibnamefont {Ness}},\ }\href@noop
  {} {\bibfield  {journal} {\bibinfo  {journal} {Journal of Fluid Mechanics}\
  }\textbf {\bibinfo {volume} {984}},\ \bibinfo {pages} {A67} (\bibinfo {year}
  {2024})}\BibitemShut {NoStop}%
\bibitem [{\citenamefont {Desmond}\ and\ \citenamefont
  {Weeks}(2014)}]{Desmond_2014}%
  \BibitemOpen
  \bibfield  {author} {\bibinfo {author} {\bibfnamefont {K.~W.}\ \bibnamefont
  {Desmond}}\ and\ \bibinfo {author} {\bibfnamefont {E.~R.}\ \bibnamefont
  {Weeks}},\ }\href@noop {} {\bibfield  {journal} {\bibinfo  {journal}
  {Physical Review E}\ }\textbf {\bibinfo {volume} {90}},\ \bibinfo {pages}
  {022204} (\bibinfo {year} {2014})}\BibitemShut {NoStop}%
\bibitem [{\citenamefont {Ogarko}\ and\ \citenamefont
  {Luding}(2012)}]{Ogarko_2012}%
  \BibitemOpen
  \bibfield  {author} {\bibinfo {author} {\bibfnamefont {V.}~\bibnamefont
  {Ogarko}}\ and\ \bibinfo {author} {\bibfnamefont {S.}~\bibnamefont
  {Luding}},\ }\href@noop {} {\bibfield  {journal} {\bibinfo  {journal} {The
  Journal of chemical physics}\ }\textbf {\bibinfo {volume} {136}} (\bibinfo
  {year} {2012})}\BibitemShut {NoStop}%
\bibitem [{\citenamefont {Breard}\ \emph {et~al.}(2023)\citenamefont {Breard},
  \citenamefont {Dufek}, \citenamefont {Charbonnier}, \citenamefont
  {Gueugneau}, \citenamefont {Giachetti},\ and\ \citenamefont
  {Walsh}}]{breard2023fragmentation}%
  \BibitemOpen
  \bibfield  {author} {\bibinfo {author} {\bibfnamefont {E.~C.}\ \bibnamefont
  {Breard}}, \bibinfo {author} {\bibfnamefont {J.}~\bibnamefont {Dufek}},
  \bibinfo {author} {\bibfnamefont {S.}~\bibnamefont {Charbonnier}}, \bibinfo
  {author} {\bibfnamefont {V.}~\bibnamefont {Gueugneau}}, \bibinfo {author}
  {\bibfnamefont {T.}~\bibnamefont {Giachetti}}, \ and\ \bibinfo {author}
  {\bibfnamefont {B.}~\bibnamefont {Walsh}},\ }\href@noop {} {\bibfield
  {journal} {\bibinfo  {journal} {Nature Communications}\ }\textbf {\bibinfo
  {volume} {14}},\ \bibinfo {pages} {2079} (\bibinfo {year}
  {2023})}\BibitemShut {NoStop}%
\end{thebibliography}%
\bibliographystyle{apsrev4-1}
\end{document}